\documentclass[  superscriptaddress, amsmath,amssymb, aps,twocolumn,longbibliography]{revtex4-1} %longbibliography
\usepackage{graphicx}% Include figure files
\usepackage{dcolumn}% Align table columns on decimal point
\usepackage{amsthm}
\usepackage{bm}% bold math

\pdfoutput=1
\usepackage{graphicx,graphics,epsfig,subfigure,times,bm,bbm,amssymb,amsmath,amsthm,mathrsfs,MnSymbol}
\usepackage{gensymb}
\usepackage{amsfonts}
\usepackage[matrix,frame,arrow]{xypic}
\usepackage[pdfstartview=FitH]{hyperref}
\hypersetup{
    colorlinks=true,       % false: boxed links; true: colored links
    linkcolor=red,          % color of internal links
    citecolor=magenta,        % color of links to bibliography
    filecolor=magenta,      % color of file links
    urlcolor=cyan,           % color of external links
    runcolor=cyan}
\usepackage{epstopdf}
\usepackage[pdftex]{color}
\usepackage{braket}%Dirac Notation in QM
\usepackage{enumerate}
\usepackage[normalem]{ulem}
\usepackage[usenames,dvipsnames]{xcolor}
\usepackage{multirow}
\usepackage{mathtools}

\definecolor{orange}{rgb}{1,0.5,0}

\begin{document}

\title{Characterizing dynamical criticality of many-body localization transitions from a Fock-space perspective}

\author{Zheng-Hang Sun}
\affiliation{Theoretical Physics \uppercase\expandafter{\romannumeral3}, Center for Electronic Correlations and Magnetism, Institute of Physics, University of Augsburg, D-86135 Augsburg, Germany}

\author{Yong-Yi Wang}
\email{yywang@iphy.ac.cn}
\affiliation{Institute of Physics, Chinese Academy of Sciences, Beijing 100190, China}
\affiliation{School of Physical Sciences, University of Chinese Academy of Sciences, Beijing 100190, China}

\author{Jian Cui}
\email{jianCui@buaa.edu.cn}
\affiliation{School of Physics, Beihang University, Beijing 100191, China}

\author{Heng Fan}
\email{hfan@iphy.ac.cn}
\affiliation{Institute of Physics, Chinese Academy of Sciences, Beijing 100190, China}
\affiliation{School of Physical Sciences, University of Chinese Academy of Sciences, Beijing 100190, China}
\affiliation{Songshan Lake  Materials Laboratory, Dongguan 523808, Guangdong, China}
\affiliation{Beijing Academy of Quantum Information Sciences, Beijing 100193, China}
\affiliation{Hefei National Laboratory, Hefei 230088, China}

\author{Markus Heyl}
\affiliation{Theoretical Physics \uppercase\expandafter{\romannumeral3}, Center for Electronic Correlations and Magnetism, Institute of Physics, University of Augsburg, D-86135 Augsburg, Germany}

\begin{abstract}
\noindent Characterizing the nature of many-body localization transitions (MBLTs) and their potential critical behaviors has remained a challenging problem. In this work, we study the dynamics of the displacement, quantifying the spread of the radial probability distribution in the Fock space, for three systems with MBLTs, i.e., the Hamiltonian models with quasiperiodic and random fields, as well as a random-circuit Floquet model of a MBLT. We then perform a finite-size scaling analysis of the long-time averaged displacement by considering two types of ansatz for MBLTs, i.e., continuous and BKT transitions. The data collapse based on the assumption of a continuous phase transition with power-law correlation length reveals that the scaling exponent of the MBLT induced by random field is close to that of the Floquet model, but significantly differes from the quasiperiodic model. Additionally, we find that the BKT-type scaling provides a more accurate description of the MBLTs in the random model and the Floquet model, yielding larger (finite-size) critical points compared to those obtained from power-law scaling. Our work highlights that the displacement is a valuable tool for studying MBLTs, as relevant to ongoing experimental efforts.
\end{abstract}
\pacs{Valid PACS appear here}
\maketitle

\section{Introduction}

The study of non-equilibrium phases and phase transitions of many-body quantum systems is a central focus in modern physics~\cite{Eisert:2015ws}. In particular, in disordered systems a paradigmatic non-equilibrium phase transition is expected to occur from an ergodic to the many-body localized (MBL) phase upon increasing the disorder strength, known as the many-body localization transition (MBLT)~\cite{Nandkishore:2015vg,Abanin:2017vf,Altman:2018vb,Abanin:2019va,2024arXiv240307111S,PhysRevB.82.174411,PhysRevLett.115.187201,PhysRevX.5.041047,PhysRevX.5.031033,PhysRevX.5.031032,PhysRevX.7.021013,PhysRevLett.119.110604,PhysRevLett.123.180601,PhysRevLett.119.075702,PhysRevLett.121.206601,PhysRevB.91.081103,PhysRevB.107.115132,Rispoli:2019vb}. 
Although the phenomenology of the deep MBL phase based on the local integrals of  motions has made substantial progress~\cite{PhysRevB.90.174202,PhysRevLett.111.127201}, 
precisely characterizing the critical properties of MBLTs, such as the location of transition points and their universality classes, has remained an outstanding challenge~\cite{2024arXiv240307111S}. 

One general method for identifying critical properties of continuous phase transitions is to study a data collapse with a subsequent finite-size scaling analysis. In the context of MBL, this is typically based on the properties of highly-excited eigenstates and eigenvalues obtained from exact diagnolization (ED) calculations, with the level-spacing ratio as a prominent example~\cite{PhysRevB.91.081103,PhysRevB.107.115132,PhysRevB.75.155111,PhysRevB.92.195153,PhysRevB.96.075146,PhysRevB.97.220201,PhysRevB.97.201105,PhysRevB.98.094202,PhysRevB.100.134504,PhysRevB.102.064207}. However, if the ansatz with power-law correlation length is adopted for the scaling analysis, the scaling exponent $\nu$ obtained in this way usually violates some general bounds, such as the Harris-Chayes bound $\nu>2$ for the MBLT in one-dimensional (1D) systems induced by random fields~\cite{Harris:1974tn,PhysRevLett.57.2999,chandran2015finite}. The violation of the theoretical bounds motivates the studies of MBLTs employing the Berezinskii-Kosterlitz-Thouless (BKT) scaling~\cite{PhysRevB.102.064207,PhysRevE.102.062144,PhysRevResearch.2.042033,PhysRevB.104.214201}. An alternative way to study MBLTs is to focus on the dynamics of local observables, such as the imbalance, which is of key importance due to its experimental relevance~\cite{Schreiber:2015ub,Choi:2016vl,Smith:2016uo,PhysRevLett.119.260401,PhysRevX.7.041047,PhysRevResearch.4.013148,PhysRevResearch.3.033043,Guo:2021wr}. However, it has been pointed out that there is an obvious discrepancy between the transition points estimated by using the level-spacing ratio and those characterized by the freezing dynamics of imbalance, as another widely adopted landmark of MBLTs~\cite{PhysRevB.98.174202,PhysRevB.100.104203,PhysRevB.101.035148,PhysRevB.105.224203}. 

Recent developments highlight the suitability to characterize MBLTs upon studying the quantum states in the Fock space~\cite{PhysRevB.92.134204,PhysRevLett.115.046603,PhysRevB.99.045131,PhysRevB.99.165131,PhysRevB.101.134202,PhysRevB.106.054203,PhysRevB.104.174201,PhysRevB.104.024202,2024arXiv240103027G,PhysRevB.108.L140201}. In particular, this includes the so-called radial distribution measuring the probability $\Pi(x)$ for an initially localized wave packet in Fock space to delocalize over a distance $x$~\cite{PhysRevB.104.174201,PhysRevB.104.024202}. 
Importantly, it has been shown that the distribution becomes distinctively broad in the vicinity of the MBLTs~\cite{PhysRevB.104.174201,PhysRevB.104.024202}. In addition to the theoretical studies, it is worthwhile to note a recent experiment where the Fock-space dynamics in a disordered superconducting circuit has been measured, showing that the radial distribution can be experimentally feasible probes for signaling MBLTs~\cite{Yao:2023tc}. Although there has been both theoretical and experimental progress in understanding MBLTs from the Fock-space perspective, a detailed and comprehensive finite-size scaling analysis, which plays a key role in characterizing precise critical properties of quantum phase transitions~\cite{Cardy_1996}, of the radial distribution has remained absent so far.

In this work, we show that the width of the radial probability distribution, the so-called displacement, turns out to be a suitable quantity to characterize MBLTs through data collapse and finite-size scaling analysis. We reveal that the critical behaviors of the MBLT for the quasiperiodic model, compellingly described by the power-law scaling, are distinct from those in the random Hamiltonian model and the random-circuit Floquet model, which are more promisingly described by the BKT scaling. More importantly, the BKT scaling based on the displacement gives a larger estimation of (finite-size) critical points than conventional probes of MBLTs, such as the level spacing ratio~\cite{PhysRevB.102.064207}. By using the displacement, the results of the BKT scaling are consistent with the analysis based on the many-body resonances~\cite{PhysRevB.105.174205}. Our work opens the door to also experimentally identify the dynamical phase diagram and universality classes of MBLTs using large-scale analog quantum simulations with the potential to overcome the numerical limitations from finite-size effects.

\section{Models and probes}

We focus on three models including the 1D spin-$1/2$ Heisenberg model with random as well as quasiperiodic field, and a Floquet random-circuit model for the MBLT~\cite{PhysRevB.98.134204,PhysRevB.105.174205}. Let us first introduce the considered Heisenberg model: 
\begin{eqnarray}
\hat{H} = \hat{H}_{XY} + \hat{H}_{ZZ} + \hat{H}_{Z}, 
\label{H_all}
\end{eqnarray}
where $\hat{H}_{XY} = J\sum_{r=1,2} \sum_{j=1}^{L-r}(\hat{S}^{x}_{j}\hat{S}^{x}_{j+r} + \hat{S}^{y}_{j}\hat{S}^{y}_{j+r})$, $\hat{H}_{ZZ} =J \sum_{j=1}^{L-1}\hat{S}^{z}_{j}\hat{S}^{z}_{j+1}$, with $J$ being the coupling strength, and $\hat{H}_{Z} = W \sum_{j=1}^{L}\cos(2\pi k j + \phi_{j}) \hat{S}_{j}^{z}$ describing the on-site disordered fields with the strength $W$, and $k = (\sqrt{5}-1)/2$. For the quasiperiodic and random model, we consider the global phase offset $\phi_{j} = \phi \in [0,2\pi)$ for all $j=1,2,..., L$, and the $\phi_{j}$ randomly drawn from a uniform distribution $[0,2\pi)$ , respectively. For the Hamiltonian (\ref{H_all}), the location of critical points estimated by the ED calculations of the level spacing ratio is $W^{(r)} \simeq 4.3$ and $5.5$ for the quasiperiodic and random model, respectively~\cite{PhysRevLett.119.075702}. Moreover, both the numerical works based on the ED~\cite{PhysRevLett.119.075702} and the renormalization group~\cite{PhysRevLett.121.206601} indicate that the MBLTs in random and quasiperiodic systems belong to two different universality classes. 

In addition to Heisenberg models (\ref{H_all}), whose dynamics has both the conservation of energy and $U(1)$ symmetry, we consider a Floquet model~\cite{PhysRevB.105.174205}
\begin{eqnarray}
\hat{U}_{F} = \hat{U}_{u}\hat{U}_{d},
\label{uf}
\end{eqnarray}
where $\hat{U}_{d} = \bigotimes_{i=1}^{L} \hat{d}_{i}$
with $\hat{d}_{i}$ being a single-qubit gate at $i$-th site, generated by sampling a $2\times 2$ random matrix from the circular unitary ensemble and then diagonalizing it, and $\hat{U}_{u} = \bigotimes_{j=1}^{L-1} \exp[(i/W)\hat{M}_{k_{j},k_{j}+1}]$
with $k_{j}\in S_{L-1}$ being a random permutation, $\hat{M}_{k_{j},k_{j}+1}$ as a $4\times 4$ random matrix sampled from the Gaussian unitary ensemble, and $W$ as the effective disorder strength. The Floquet model is free of any conservation laws, eliminating the influence of conserved quantities on the physics of MBLTs.
Recent work shows that the critical point of the MBLT in the Floquet model (\ref{uf}) estimated by the level spacing ratio is $W^{(r)}\simeq 6$, while based on the resonance properties, the critical point satisfies $W_{c}>13$~\cite{PhysRevB.105.174205}.

We now introduce the radial probability distribution defined in the Fock space. Consider a quantum many-body state  $|\psi \rangle$, which can be chosen as a highly-excited eigenstate of a disordered model~\cite{PhysRevB.104.174201,PhysRevB.104.024202}, or a non-equilibrium state generated through  time evolution~\cite{PhysRevB.100.214313,PhysRevB.107.094206}.
Then, the radial probability distribution $\Pi(x)$ can be defined as $\Pi(x) = \sum_{D_{Z,Z^{*}} = x}  |\langle Z  | \psi \rangle |^{2}$, where $| Z\rangle$ represents the product states in $z$-basis, i.e., $| Z\rangle = \otimes_{j=1}^{L}|z_{j}\rangle $ with $z_{j} \in \{ -1,+1\} $ corresponding to the eigenstate of $\sigma_{j}^{z}$ with the eigenvalue $-1$ and $+1$ [see Supplementary Materials for more details of $\Pi(x)$].
Here, $D_{Z,Z^{*}}$ refers to the Hamming distance between the state $|Z\rangle$ and $|Z^{*}\rangle$, where $|Z^{*}\rangle$ labels the closest Fock state to 
$| \psi \rangle $, i.e.,  obtained through 
$\max_{Z}  |\langle Z  | \psi \rangle |^{2} = |\langle Z^{*}  | \psi \rangle |^{2}$. The spread of the radial probability distribution can be characterized by defining the moments $\widetilde{X^{n}} = \sum_{x} x^{n}  \Pi(x)$, 
and the displacement 
\begin{eqnarray}
\Delta X^{2} =   \widetilde{X^{2}}  - (\widetilde{X})^{2}.  
\label{rpd_2}
\end{eqnarray}
The wave packet dynamics in the Fock space has been widely employed to study the single-particle AL transitions~\cite{doi:10.1143/JPSJ.66.314,PhysRevLett.79.1959,PhysRevE.108.054127}, and recently extended to many-body systems~\cite{Yao:2023tc}.

For what follows we will adopt a non-equilibrium dynamical setting in close analogy to the capabilities in analog quantum simulation devices~\cite{Yao:2023tc}.
Concretely, we study the time-dependent state $|\psi(t)\rangle = \exp (-i\hat{H}t)|Z_{0}\rangle$ for the Hamiltonian systems whereas $|\psi(n)\rangle = \hat{U}_{F}^{n}|Z_{0}\rangle$ for the Floquet model.
As initial state we choose  the N\'eel state $|Z_{0}\rangle = \otimes_{j=1}^{L}|z_{j}\rangle $ with  $z_{j} = (-1)^{j+1}$.
In order to characterize the dynamics we mainly focus on the displacement $\Delta X^{2}$ in what follows.
We simulate the Hamiltonian dynamics by using Krylov subspace methods~\cite{https://doi.org/10.1002/andp.201600350}, whereas the Floquet model $\hat{U}_{F}$ is solved by employing exact numerical simulation. 
We employ at least 300 realizations for the disorder averaging for each system size $L$. 

In Appendix A, we show an example for the detailed properties of $\Pi(x)$ for non-equilibrium states with different disorder strengths. 

\begin{figure}[]
	\centering
	\includegraphics[width=1\linewidth]{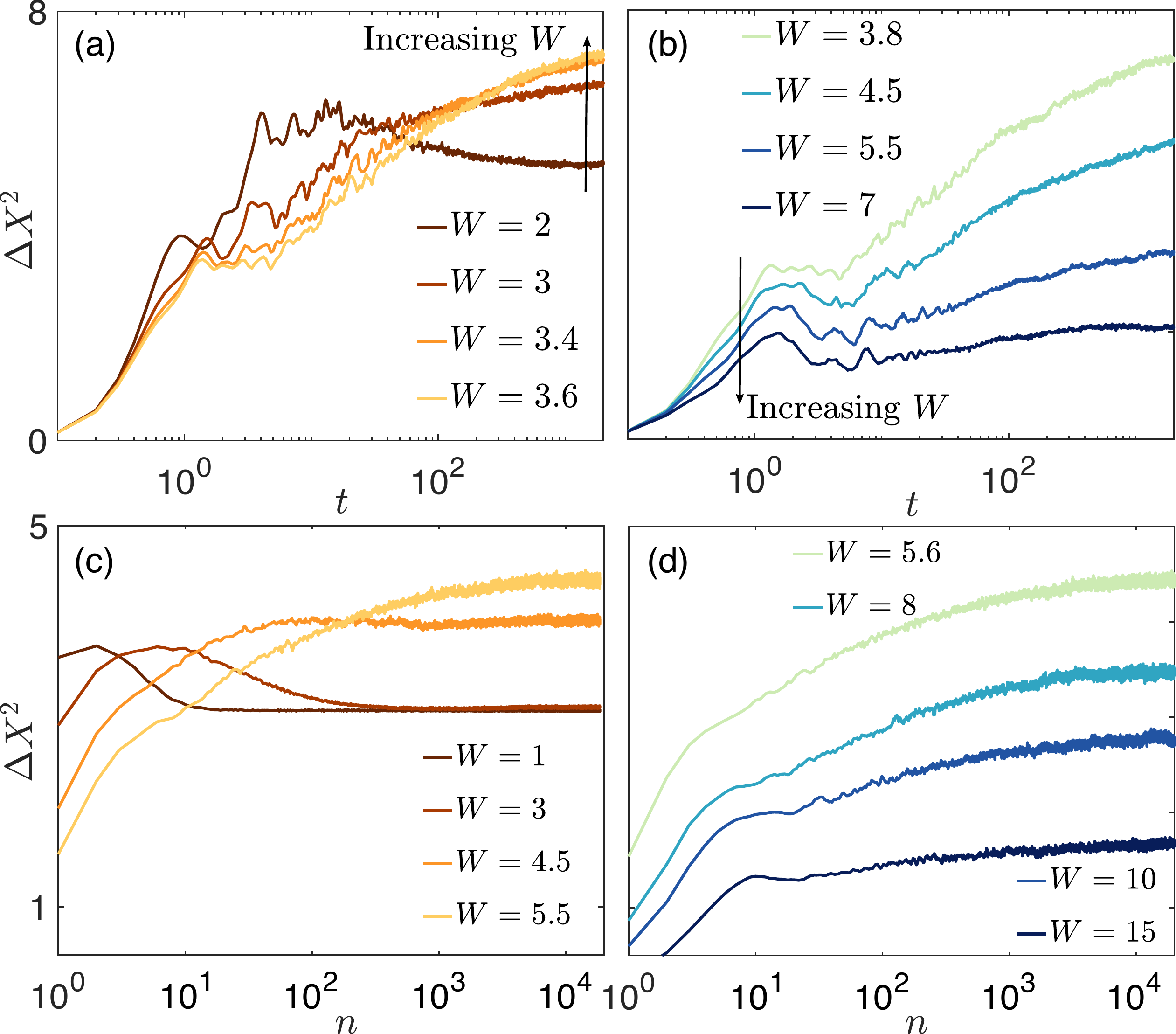}\\
	\caption{ (a) Time evolution of the displacement $\Delta X^{2}$ for the Hamiltonian (\ref{H_all}) with quasiperiodic fields and different disorder strengths $W$. (b) is similar to (a) but for larger $W$.  In (a) and (b), the time $t$ is in units where the coupling strength $J=1$. (c) is similar to (a) but for the Floquet model (\ref{uf}). (d) is similar to (c) but for larger $W$. }\label{fig1}
\end{figure}

\begin{figure}[]
	\centering
	\includegraphics[width=1\linewidth]{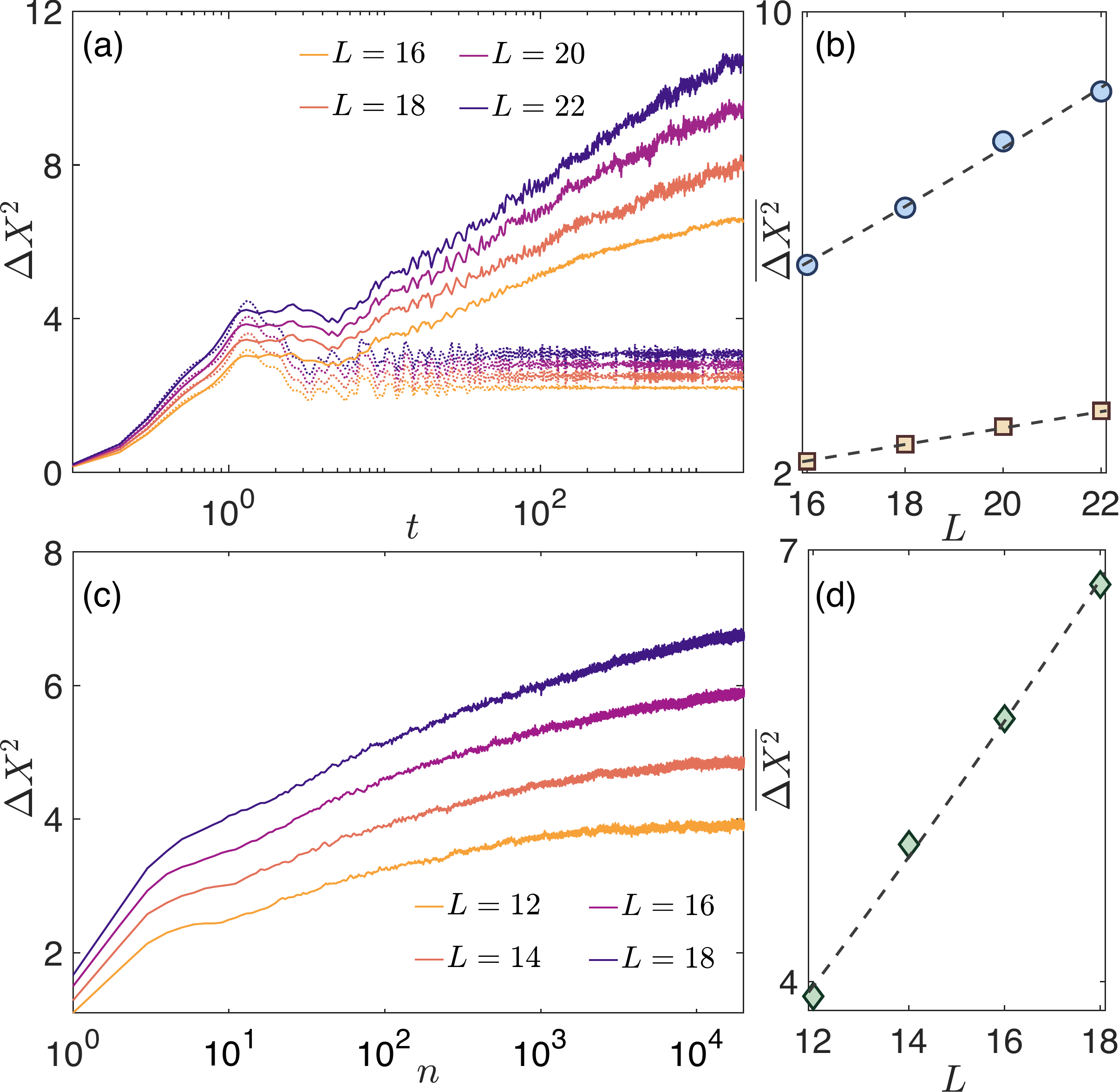}\\
	\caption{(a) The solid (dotted) lines represent the time evolution of $\Delta X^{2}$ for the Hamiltonian (\ref{H_all}) (non-interacting model $\hat{H}'$) with a strength $W=4.1$ and different system sizes. The time $t$ is in units where the coupling strength $J=1$. (b)  The time-averaged $\Delta X^{2}$, i.e., $\overline{\Delta X^{2}}$,  as a function of $L$ with the time interval $t\in[10,1000]$. The circle and square data correspond to the interacting model with the Hamiltonian (\ref{H_all}) and the non-interacting one $\hat{H}'$, respectively. (c) The time-evolution of $\Delta X^{2}$ for the Floquet model (\ref{uf}) with the disorder strength $W=7$ and different system sizes. (d) The  $\overline{\Delta X^{2}}$ as a function of  $L$ for the Floquet model with the time interval $n\in[10^{4},2\times 10^{4}]$. The dashed lines in (b) and (d) show the power-law fittings $\Delta X^{2} \propto L^{\beta}$. }\label{fig1_add}
\end{figure}

\begin{figure*}[]
	\centering
	\includegraphics[width=1\linewidth]{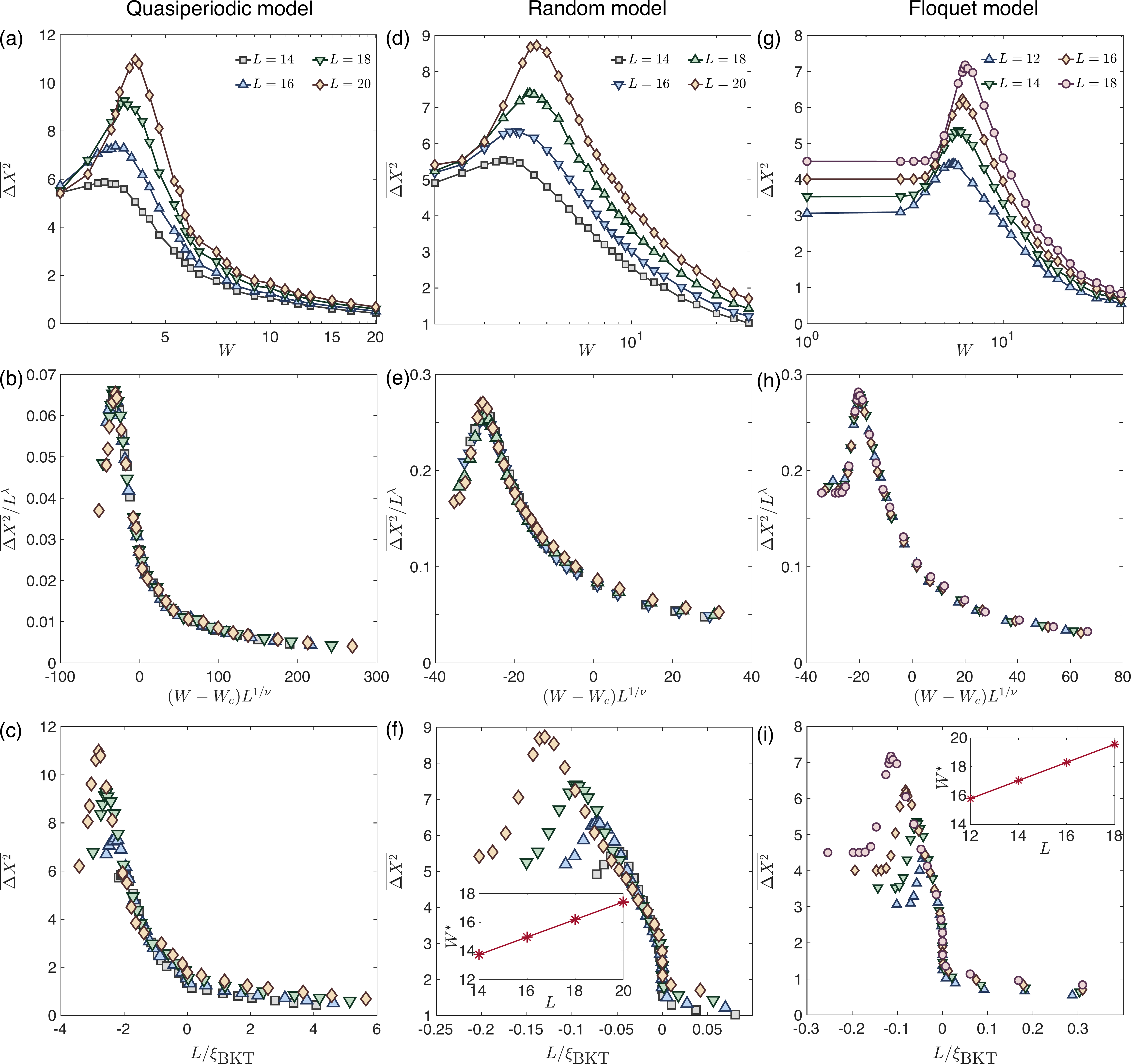}\\
	\caption{(a) The value of  $\overline{\Delta X^{2}}$ as a function of disorder strength $W$, for the quasiperiodic model with sizes $L=14, 16, 18$ and $20$. (b)  Finite-size data collapse for the data of $\overline{\Delta X^{2}}$ shown in (a) based on the power-law scaling, with $W_{c} = 5.7\pm 0.4$, $\nu=1.0\pm0.1$, and $\lambda = 1.7\pm 0.1$. (c) Finite-size data collapse for the data of $\overline{\Delta X^{2}}$ shown in (a) based on the BKT scaling, with $b=4.2\pm 0.2$, $w_{0} = 6.8 \pm 0.5$, and $w_{1} = 0.1\pm 0.005$. (d) is similar to (a) but for the random model. (e) Finite-size data collapse for the data in (c) based on the power-law scaling, with $W_{c} = 14.7\pm1.8$, $\nu=2.9\pm0.3$, and $\lambda = 1.2\pm0.2$. (f) is similar to (e) but for the BKT scaling, with $b=18\pm 0.9$, $w_{0} = 5.2 \pm 0.2$, and $w_{1} = 0.61\pm 0.03$. (g) The value of  $\overline{\Delta X^{2}}$ as a function of $W$, for the Floquet model with sizes $L=12, 14, 16$ and $18$. (h) Finite-size data collapse for the data in (g) based on the power-law scaling, with $W_{c} = 14.3\pm1.7$, $\nu=3.0\pm0.7$, and $\lambda = 1.1\pm0.1$. (i) is similar to (h) but for the BKT scaling, with $b=18.4\pm 1.0$, $w_{0} = 8.2 \pm 0.7$, and $w_{1} = 0.63\pm 0.03$. The insets of (f) and (i) show the function $W^{*} = w_{0} + w_{1}L$.  }\label{fig2_new}
\end{figure*}

\section{Dynamics}

We first consider the Heisenberg model in Eq.~(\ref{H_all}) with quasiperiodic fields. The results for different strengths of the field $W$ at a system of size $L=16$ are plotted in Fig.~\ref{fig1}(a) and (b). In the ergodic phase ($W=2$), after an initial increase and some intermediate drop, as a non-monotonic behavior, the displacement $\Delta X^{2}$ finally approaches to a static value on the displayed time scales. For larger values of the disorder $W>2$, $\Delta X^{2}$ monotonically increases during the time evolution. More importantly, according to the results in Fig.~\ref{fig1}(a) and (b), it is seen that the late-time value of $\Delta X^{2}$  
exhibits a peak around $W^{*} \simeq 3.6$ for the system of size $L=16$, which is similar to the properties of the displacement $\Delta X^{2}$ for highly-excited eigenstates~\cite{PhysRevB.104.174201,PhysRevB.104.024202}. We also simulate the dynamics of $\Delta X^{2}$ for the Floquet model in Eq.~(\ref{uf}). As shown in Fig.~\ref{fig1}(c) and (d), the dynamical behaviors of $\Delta X^{2}$ are similar to those in the Heisenberg model (\ref{H_all}).

Next, we determine the dynamics of $\Delta X^{2}$ for the Heisenberg model (\ref{H_all}) at different system sizes and an intermediate quasiperiodic-field strength $W=4.1$. In order to highlight the effect of  interactions  we also consider a non-interacting model $\hat{H}' =   \sum_{j=1}^{L-1}(\hat{S}^{x}_{j}\hat{S}^{x}_{j+1} + \hat{S}^{y}_{j}\hat{S}^{y}_{j+1})  + \hat{H}_{Z}$ with quasiperiodic fields, i.e., $\phi_{j} = \phi \in [0,2\pi)$ in $H_{Z}$, where a transition between the extended phase and Anderson localization (AL) occurs at $W_{c} = 2$~\cite{aubry1980analyticity}. As shown in Fig.~\ref{fig1_add}(a), the dynamical behaviors of $\Delta X^{2}$ for the interacting system are distinct from the non-interacting AL. For the AL, after a short-time increase, $\Delta X^{2}$ saturates to relatively small values. In contrast, for the interacting case, there is an approximately logarithmic growth of the $\Delta X^{2}$. Theoretical studies indicate that near the critical point of the MBLT, $\Delta X^{2}$ satisfies $\Delta X^{2} \sim L^{\beta}$ with $1<\beta \leq 2$~\cite{PhysRevB.104.174201,PhysRevB.104.024202}. By fitting the time-averaged $\Delta X^{2}$, defined as $\overline{\Delta X^{2}} = [\int_{t_{i}}^{t_{f}} \text{d}t \Delta X^{2}(t)](t_{f}-t_{i})$ with the time interval $t\in[t_{i},t_{f}]$, in Fig.~\ref{fig1_add}(b), we show that $\overline{\Delta X^{2}} \sim L^{1.36}$ for the interacting case, while for the AL,  $\overline{\Delta X^{2}} \sim L^{1.06}$ with an approximately linear size dependence. Moreover, for the Floquet model (\ref{uf}), an approximately logarithmic growth of the $\Delta X^{2}$ can also be observed  [see Fig.~\ref{fig1_add}(c)]. In Fig.~\ref{fig1_add}(d), we plot the time-averaged $\Delta X^{2}$ as a function of system size $L$, showing that $\overline{\Delta X^{2}} \sim L^{1.32}$. We note that the details of the logarithmic fittings for the dynamics of $\Delta X^{2}$, and the power-law fittings for  $\overline{\Delta X^{2}}$ as a function of $L$ in Fig.~\ref{fig1_add} are presented in the Appendix B.

\section{Scaling analysis}

After studying the temporal properties of $\Delta X^{2}$, we now focus on a scaling analysis of the time-averaged values of $\Delta X^{2}$, denoted as $\overline{\Delta X^{2}}$. To address the finite-time effect, for the quasiperiodic and random models, we consider a long-time interval $t\in[t_{H},t_{H}+100]$, with $t_{H}$ denoting the Heisenberg time, for the average (see Appendix C for more details). For the random-circuit Floquet model, we analyze the finite-time effect in Appendix C, and chose the time scale $n\sim 10^{4}$ for the average, where the finite-time effect becomes moderate. Specifically, for system sizes $L=12$, $14$ and $16$, the time interval is $n\in[10^{4},2\times 10^{4}]$, and for the system size $L=18$, $n\in[3\times 10^{4}, 4\times 10^{4}]$.

We plot the $\overline{\Delta X^{2}}$ for different sizes $L$ of all three models, i.e., the quasiperiodic model, random model, and Floquet model, in Fig.~\ref{fig2_new}(a), (d), and (g), respectively. One can see that with increasing $L$, the location of the peak $W^{*}$ shifts to larger value.

To more accurately capture the critical behaviors of MBLTs in the three models, we perform the data collapse of $\overline{\Delta X^{2}}$ as a function of disorder strength $W$ based on both the scaling function with the power-law and BKT-type correlation lengths. We first consider the ansatz of continuous phase transitions with a power-law scaling function
\begin{eqnarray}
\overline{\Delta X^{2}} (L,W) = L^{\lambda} g[(W-W_{c})L^{1/\nu}],
\label{scaling}
\end{eqnarray}
which is based on the assumption of a divergent  correlation length. The power-law scaling (\ref{scaling}) can efficiently capture the critical properties of AL transitions in high-dimensional systems~\cite{PhysRevLett.108.095701,Slevin_2014,PhysRevResearch.2.042031}, and is also widely employed in the study of  MBLTs~\cite{PhysRevLett.119.075702,PhysRevLett.121.206601,PhysRevB.91.081103,PhysRevB.107.115132}. 

We also focus on the BKT-type scaling function $\overline{\Delta X^{2}} (L,W) = L/\xi_{\text{BKT}}$, with the BKT-type correlation length 
\begin{eqnarray}
\xi_{\text{BKT}} = \exp \left\{\frac{b}{\sqrt{|W-W^{*}|}} \right\},
\label{bkt}
\end{eqnarray}
where the crossing point ansatz is $W^{*} = w_{0} + w_{1}L$.

As shown in Fig.~\ref{fig2_new}(b), for the quasiperiodic model, the best data collapse with power-law correlation length is obtained when $W_{c} = 5.7\pm 0.4$, $\nu=1.0\pm0.1$, and $\lambda = 1.7\pm 0.1$. The scaling exponent $\nu$ satisfies the Harris-Luck bound $\nu\geq 1$ for 1D quasiperiodic systems~\cite{Luck:1993vn}. In comparison with the result of BKT-type scaling shown in Fig.~\ref{fig2_new}(c), we suggest that power-law scaling describes better for the MBLT induced by the quasiperiodic field. 

For the random model, as displayed in Fig.~\ref{fig2_new}(e) and (f), although the best power-law scaling with $W_{c} = 11.7\pm 2.0$, $\nu = 3.8\pm 0.5$, and $\lambda = 1.1\pm 0.1$ gives a scaling exponent satisfying the Harris-Chayes bound $\nu\geq 2$~\cite{Harris:1974tn,PhysRevLett.57.2999,chandran2015finite}, based on the comparison between the $\overline{\Delta X^{2}}$ after the data collapse in the vicinity of critical points, i.e., $(W-W_{c})L^{1/\nu} \sim 0$ for the power-law scaling, and $L/\xi_{\text{BKT}}\sim 0$ for the BKT-type scaling, one can see that the BKT-type scaling more promisingly describes the MBLT in the random model. Similar signatures can also be observed in the Floquet model [see Fig.~\ref{fig2_new}(h) and (i)]. In short, the BKT-type scaling provides a better data collapse close to the critical point for both the random Hamiltonian model and the random-circuit Floquet model.   

We also characterize the accuracy of the obtained data collapse based on the power-law correlation length by self-consistently checking the finite-size dependence of the $\overline{\Delta X^{2}}$ at the estimated critical point. The results are presented in Appendix E, indicating that  for the random model and the Floquet model, the power-law scaling is less compelling than that of the quasiperiodic model. 

Previously, based on conventional probes of MBLTs, such as the level-spacing ratio, the BKT scaling analysis shows that the estimated (finite-size) critical points are lower than those reported in the works where the power-law scaling is employed to estimate the critical points~\cite{PhysRevB.102.064207}. However, taking the new probe $\Delta X^{2}$ into consideration, we reveal that the BKT-type scaling gives larger (finite-size) critical points [see the inset of Fig.~\ref{fig2_new}(f) and (i)]. Specifically, for the Floquet model, based on the BKT scaling, the (finite-size) critical point with $L=18$ is $W^{*}\simeq 20$, which is consistent with the critical point estimated by the analysis of many-body resonances~\cite{PhysRevB.105.174205}.

\section{Summary and discussions}

In our work we show that the critical properties of MBLTs can be characterized through a data collapse of the displacement $\Delta X^{2}$ and a subsequent finite-size scaling analysis.
Our results suggest that critical properties of the MBLT induced by the quasiperiodic fields can be well captured by the power-law scaling, while for the MBLTs in the model with random fields and the Floquet model, the BKT scaling is more compelling.

For the three studied models with MBLTs, the difference between the critical properties of the MBLT in the quasiperiodic model and those in other models may be due to the fact that the avalanche instability induced by rare thermal regions is absent for the quasiperiodic model~\cite{PhysRevB.95.155129,PhysRevLett.119.150602,PhysRevLett.121.140601,PhysRevResearch.2.033262,Leonard:2023vv}. 

We also emphasize that for the quasiperiodic model, although the power-law scaling appears more compelling with the accessible system sizes, the results in Fig.~\ref{fig2_new}(b) and (c) cannot completely rule out the BKT-type scaling in this model. An important further direction is to explore the alternative formulation of the function $W^{*} = W^{*}(L)$ in Eq.~(\ref{bkt}) for the MBLT in the quasiperiodic model.

Given that the non-equilibrium dynamics of $\Delta X^{2}$ can be efficiently measured in analog quantum simulations~\cite{Yao:2023tc}, a future perspective of our work is to experimentally study MBLTs by using the data collapse and scaling analysis in  large-scale quantum simulators, which can overcome the finite-size effects of our present numerics. Additionally, for classical simulations, with further increases computational resources and the development of algorithms, conducting a scaling analysis based on the data of displacement $\Delta X^{2}$ for systems with larger and more difference sizes is an important further task.  

Our work paves the way of studying the MBLTs in more complex systems from the Fock space point of view, such as for two-dimensional systems~\cite{Choi:2016vl,Wahl:2019wv,PhysRevLett.125.155701,PhysRevB.106.L180201}, systems with long-range interactions~\cite{PhysRevB.92.134204,PhysRevB.91.094202,PhysRevX.7.041021,PhysRevA.99.033610,PhysRevA.100.063619,PhysRevLett.125.010401,PhysRevB.101.024201}, Bose/Fermi-Hubbard models~\cite{PhysRevB.100.134504,PhysRevA.92.041601,Sierant_2018}, discrete time crystals protected by MBL~\cite{PhysRevLett.118.030401,Else:2020wn,PRXQuantum.2.030346,PhysRevB.103.224205}, and systems subjected to linear potentials exhibiting the Stark MBL~\cite{Nieuwenburg:2019vb,PhysRevLett.122.040606,PhysRevB.102.054206,PhysRevB.104.205122,Morong:2021ul}.

\begin{acknowledgments}
This work was supported by National Key Research and Development Program of China (Grant No. 2021YFA1402001), National Natural Science Foundation of China (Grants Nos. 92265207, T2121001, 11934018,12375007), Innovation Program for Quantum Science and Technology  (Grant No. 2-6), Beijing Natural Science Foundation (Grant No. Z200009), Scientific Instrument Developing Project of Chinese Academy of Sciences (Grant No. YJKYYQ20200041), the European Research Council (ERC) under the European Union’s Horizon 2020 research and innovation programme (grant agreement No. 853443), and the German Research Foundation DFG via project 499180199 (FOR 5522).

\end{acknowledgments}

\appendix

\section{Fock-space anatomy of non-equilibrium states}

\begin{figure*}[]
  \centering
  \includegraphics[width=1\linewidth]{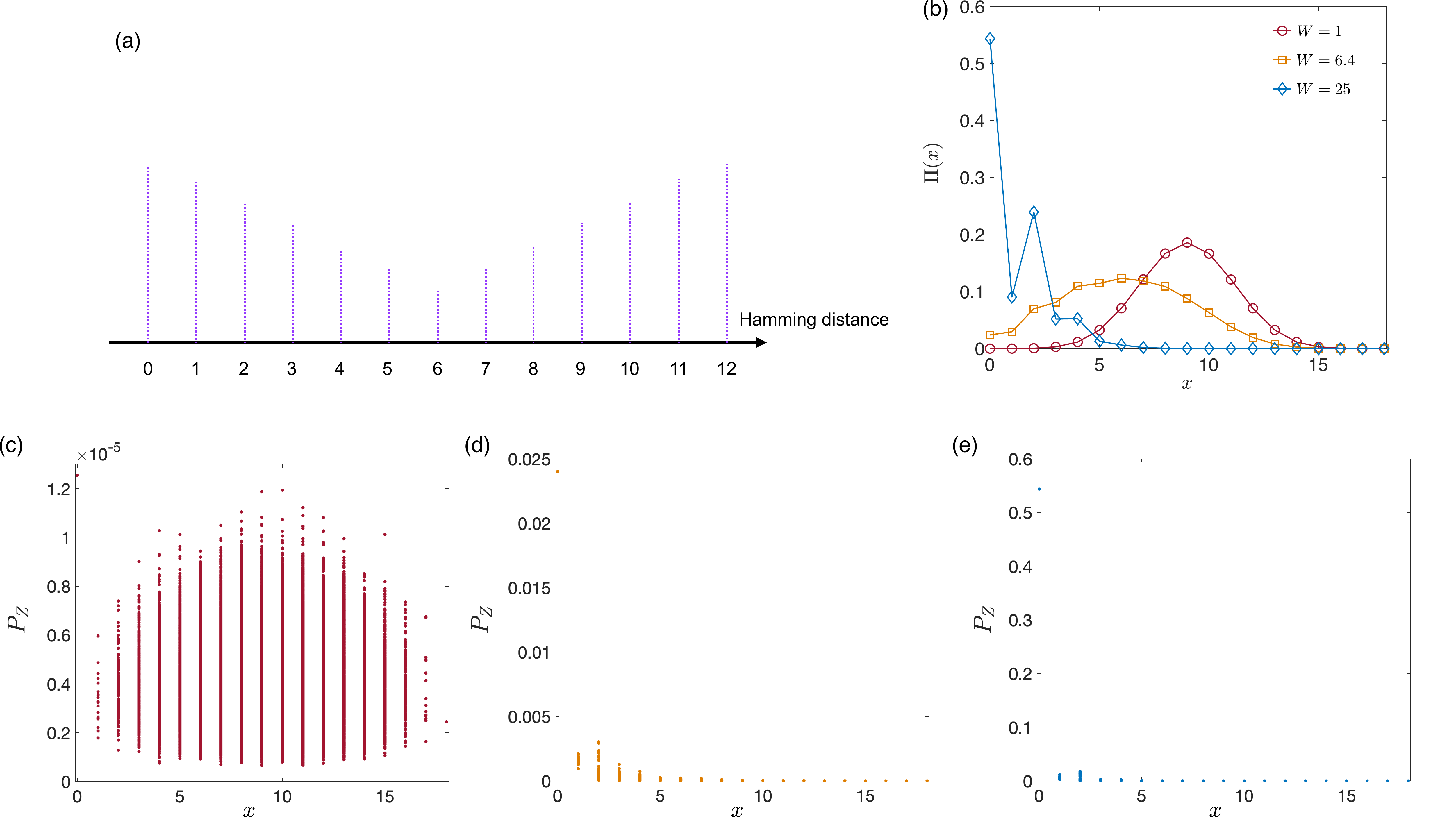}\\
  \caption{(a) A schematic of the Fock space for the Floquet model of MBL as an example with size $L=12$, where the $2^{L}$ pink nodes represent the $z$-basis vectors, classified by different Hamming distance.  (b) The radial probability distribution $\Pi(x)$ versus the Hamming distance $x$ for the non-equilibrium states $|\psi(n)\rangle$ obtained from the time evolution of the Floquet model with $n=2\times 10^{4}$ different values of $W$. (c) For the Floquet model with $W = 1$, the probabilities $P_{Z} = |\langle \psi(n)|Z\rangle|^{2}$, where $n=2\times 10^{4}$ and $|Z\rangle$ refers to the  $z$-basis vectors, versus the Hamming distance $x$ corresponding to the closest state $|Z^{*}\rangle$. (d) is similar to (c) but with $W = 6.4$. (e) is similar to (c) but with $W = 25$. }\label{fs}
\end{figure*}

In this section, we present more details of the Fock-space [see Fig.~\ref{fs}(a)] anatomy of non-equilibrium states. The first step is calculating the probabilities $P_{Z} = |\langle \psi|Z\rangle|^{2}$, with $|Z\rangle = \otimes_{j=1}^{L} |z_{j}\rangle$ ($z_{j}\in\{-1,+1\}$) being the product states in $z$-basis, and $|\psi\rangle$ being the non-equilibrium states we focus on. Next, we obtain the closest Fock state $|Z^{*}\rangle$ satisfying $\max_{Z} P_{Z} = P_{Z^{*}}$, and can calculate the Hamming distance $x$ between an arbitrary state $|Z\rangle$ and the state $|Z^{*}\rangle$, i.e., $D_{Z,Z^{*}}$. We then can plot the radial probability distribution $\Pi(x) = \sum_{D_{Z,Z^{*}}=x} P_{Z}$, see Fig.~\ref{fs}(b) as an example for the Floquet model of many-body localization (MBL) with $L=18$ and three different values of $W$. As shown in the Fig.~\ref{fig2_new}(g), with $W = 6.4$, there is a peak of the displacement $\Delta X^{2} = \sum_{x}x^{2}\Pi(x) - [\sum_{x}x\Pi(x)]^{2}$, corresponding to a broad distribution of $\Pi(x)$.  We also display the probabilities $P_{Z}$ for all the states $|Z\rangle$ in Fig.~\ref{fs}(c)-(e).

\section{Details of fitting}
\begin{figure}[]
  \centering
  \includegraphics[width=1\linewidth]{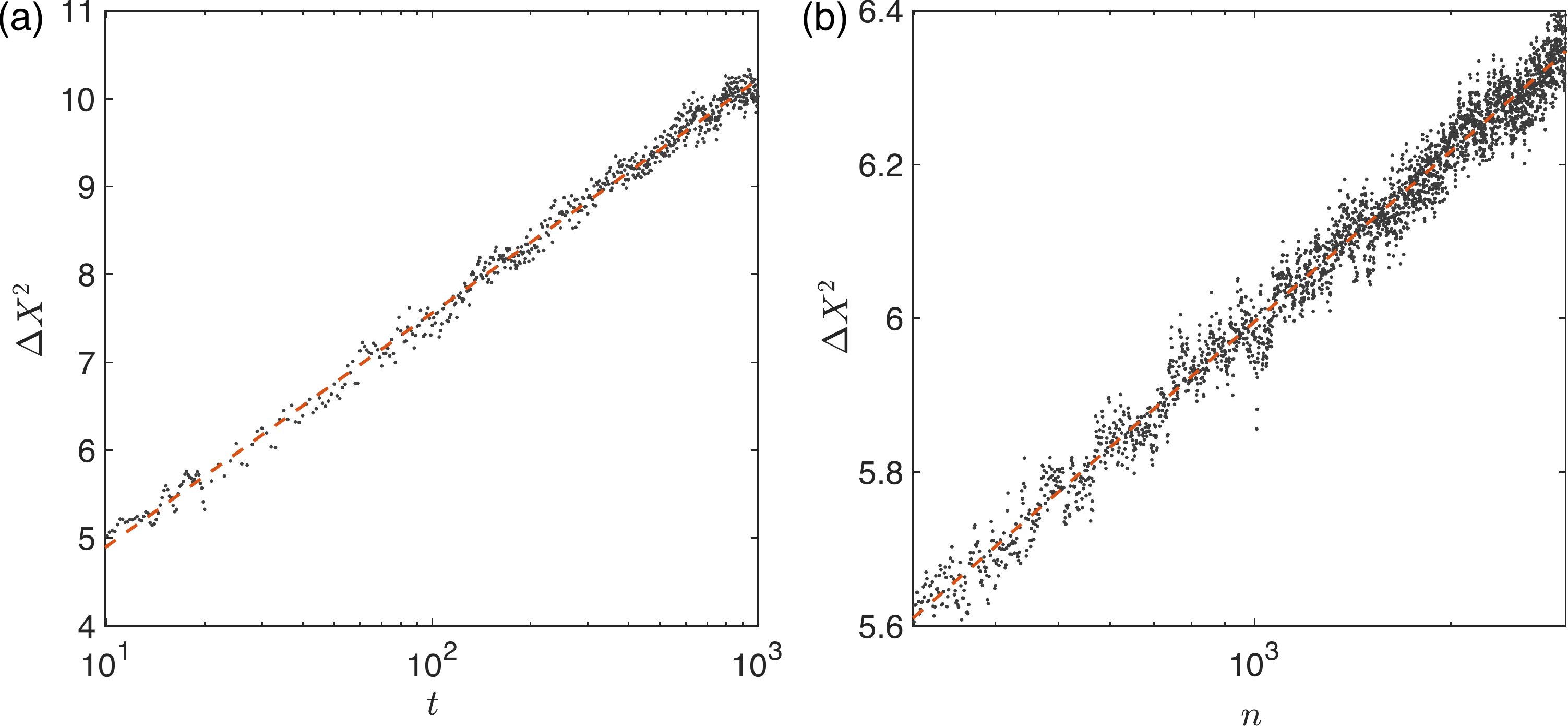}\\
  \caption{(a) The logarithmic fit of the dotted numerical data presented in Fig.~\ref{fit}(a) with $L=22$ and the time window $t\in[10,1000]$. (c)  The logarithmic fit of the dotted numerical data presented in Fig.~\ref{fit}(b) with $L=18$ and the time window $t\in[300,3000]$. The goodness of fittings are larger than $99\%$. }\label{fit}
\end{figure}

In this section, we directly present the fittings of several numerical results in the main text. In Fig.~\ref{fit}(a), we display the fitting of the numerical result shown in Fig.~\ref{fig1_add}(a) with $L=22$, taking the logarithmic function $\Delta X^{2}\propto \log t$ into consideration. For the Floquet model of MBL with $L=18$, i.e., the data in Fig.~\ref{fig1_add}(c), we also fit the dynamics of $\Delta X^{2}$ by using the logarithmic function [see Fig.~\ref{fit}(c)]. 

\section{Finite-time effect}

\begin{figure}[]
	\centering
	\includegraphics[width=1\linewidth]{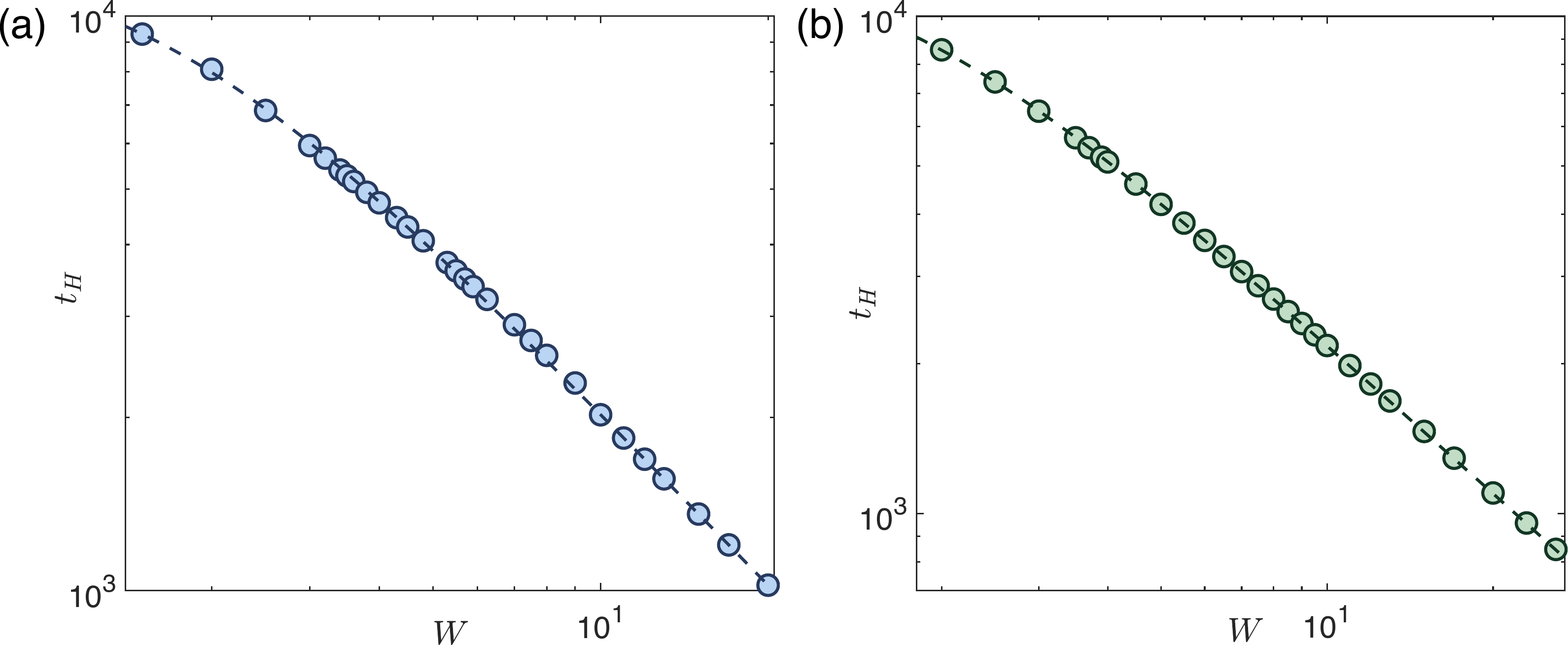}\\
	\caption{ (a) For the Hamiltonian model with quasiperiodic field and system size $L=16$, the estimated Heisenberg time $t_{H}$ as a function of the strength of disorder $W$. (b) is similar to (a) but for the Hamiltonian model with random field. The dashed line in (a) and (b) are the fitting curve with the form of function (\ref{hei_time}).  }\label{figr_tH}
\end{figure}

To perform the scaling analysis based on the dynamics of $\Delta X^{2}$, the limit of long time is required for the characterization of MBLTs. A nature cutoff of the evolved time with moderate finite-time effect is the Heisenberg time defined by $t_{H} = 2\pi/\Delta E$ with $\Delta E$ being the mean level spacing in the centre of the energy spectrum~\cite{Panda_2019}. For the Hamiltonian models considered in this work, we calculate the Heisenberg time $t_{H}$ by using the exact diagnolization. For the system size $L=16$, the results are displayed in Fig.~\ref{figr_tH}. One can see that, for both the models with the random and quasiperiodic fields, the Heisenberg time $t_{H}$ as a function of disorder strength $W$ satisfies 
\begin{eqnarray}
t_{H} = \frac{2^{L}}{L W} \frac{a}{\sqrt{b+\frac{c}{W^{2}}}},
\label{hei_time}
\end{eqnarray}
where $a$, $b$, and $c$ are fitting parameters, $L$ is the system size, and $W$ refers to the strength of disorder. For the quasiperiodic model, $a\simeq 2.857$, $b\simeq 0.3264$, and $c\simeq 0.8475$. For the random model, $a\simeq 2.112$, $b\simeq 0.1546 $, and $c\simeq 0.4062$. 

\begin{figure}[]
	\centering
	\includegraphics[width=1\linewidth]{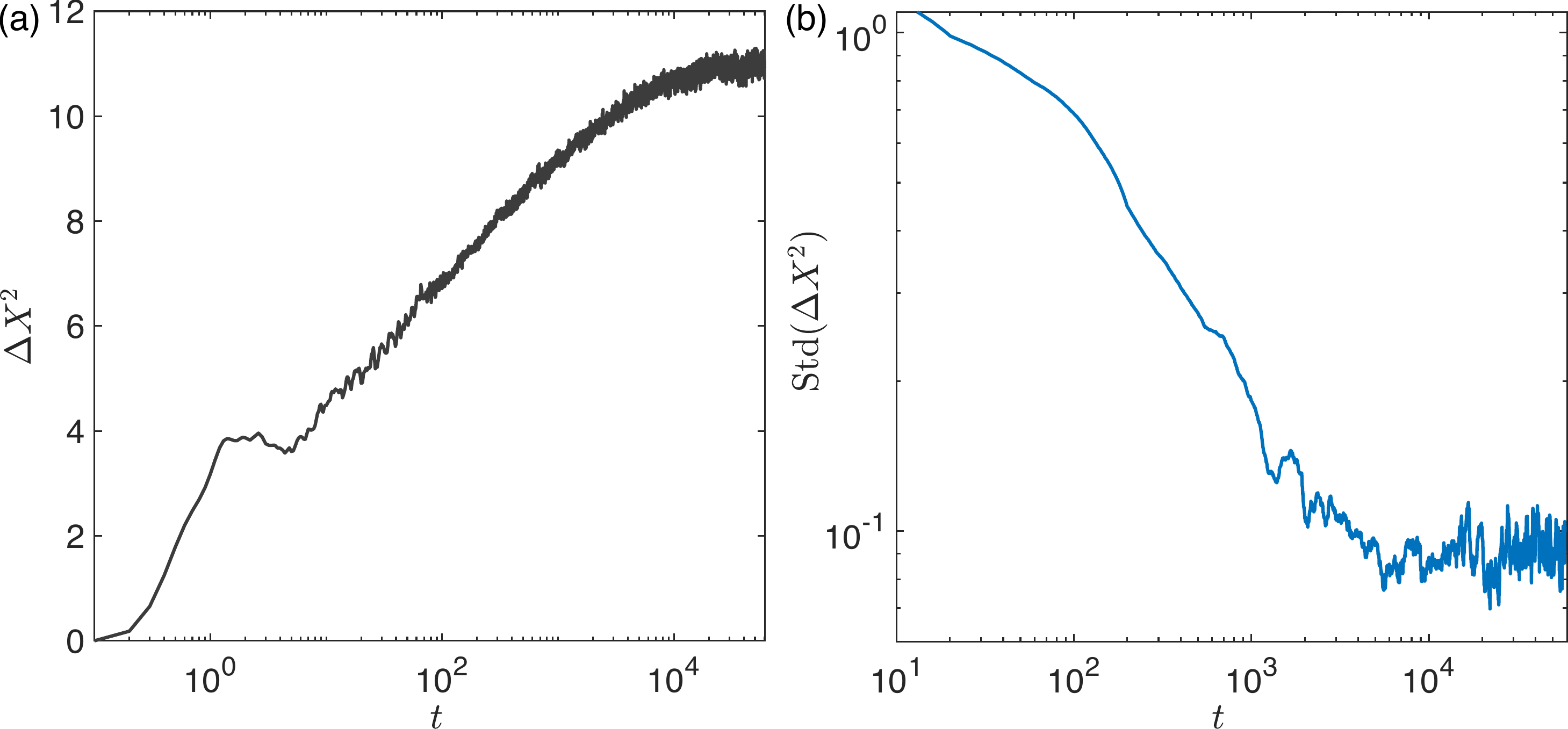}\\
	\caption{(a)  Time evolution of the displacement $\Delta X^{2}$ for the Hamiltonian defined by Eq.~(1) in the main text with quasiperiodic fields. Here, the system size is $L=20$ and the disorder strength is $W=4.1$. The finial evolved time $t\simeq 6\times 10^{4}$, which is close to the corresponding Heisenberg time. (b) The standard deviation for the data of $\Delta X^{2}$ in (a) with the time interval $[t,t+1000]$, as a function of time $t$. }\label{figr_tH_add}
\end{figure}

We then estimate whether the time interval close to $t_{H}$ is sufficiently long to suppress the finite-time effect. In Fig.~\ref{figr_tH_add}, as an example, we plot the long-time dynamics of $\Delta X^{2}$ for the quasiperiodic model with $W=4.1$ and $L=20$ up to a time $t\simeq t_{H}$, as well as the corresponding standard deviation of the data $\Delta X^{2}$, denoted as $\text{Std}(\Delta X^{2})$, with the time interval $[t,t+1000]$ as a function of time $t$.  One can see that the Heisenberg time $t_{H}$ is an appropriate time scale to have moderate finite-time effect, and we chose the time interval $t\in[t_{H},t_{H}+100]$ to perform the time average of $\Delta X^{2}$ for both the quasiperiodic and random models.

Finally, we pay attention to the finite-time effect of the random-circuit Floquet model of MBL. For this Floquet model, the Heisenberg time is equal to the Hilbert-space dimension, i.e., $n_{H} = 2^{L}$. For the systems with sizes $L=12$ and $14$, the chosen time interval $n\in[10^{4},2\times 10^{4}]$ is comparable to the Heisenberg time. However, for larger system sizes $L=16$ and $18$,  the corresponding Heisenberg time $n_{H}=65536$ and $262144$ are long time scales. Taking the numerical feasibility into consideration, we now study the temporal fluctuation of the displacement $\text{Std}(\Delta X^{2})$ in detail for the Floquet model.

\begin{figure}[]
	\centering
	\includegraphics[width=0.76\linewidth]{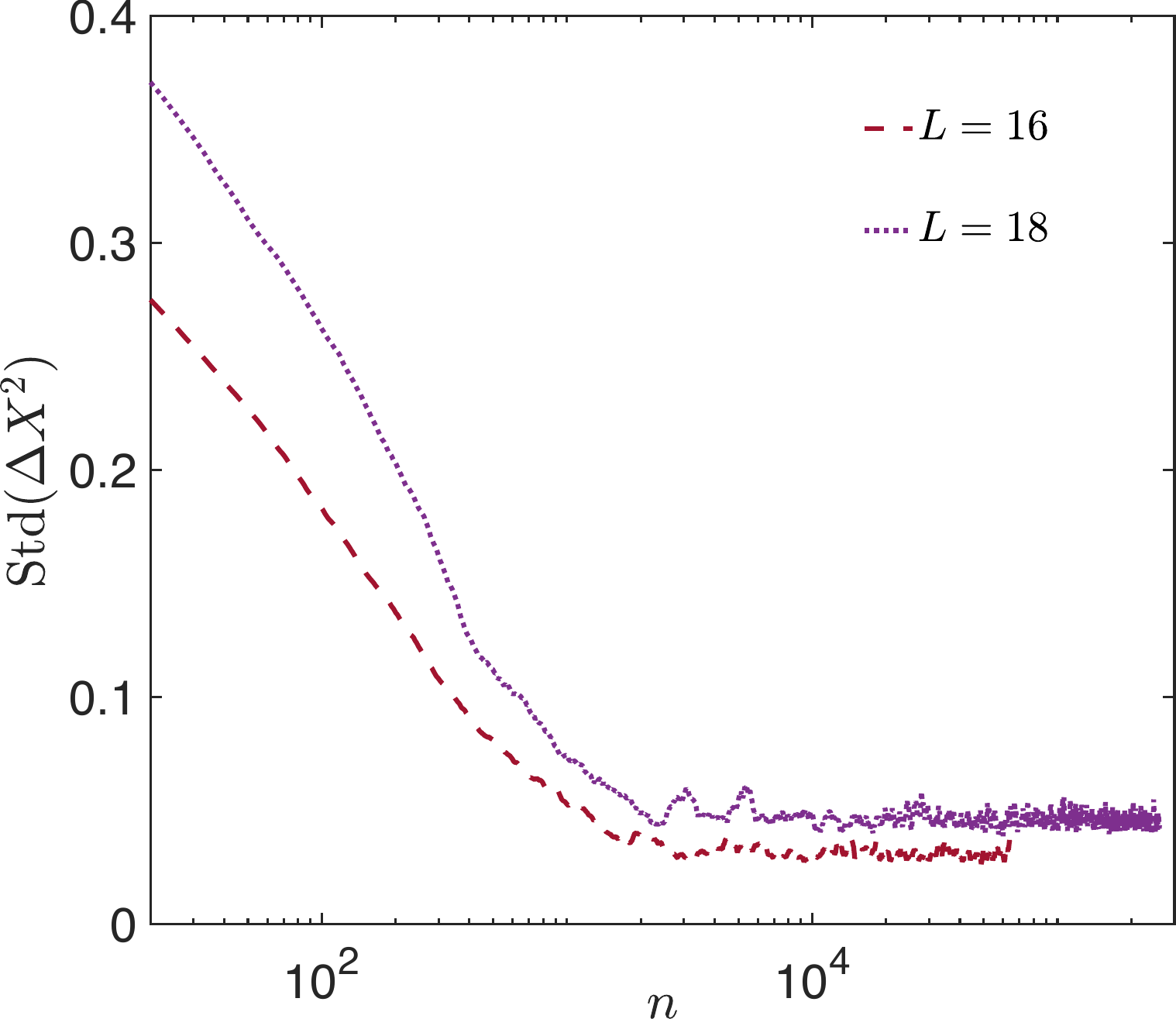}\\
	\caption{The standard deviation of the $\Delta X^{2}$ with the time interval $t\in[n,n+1000]$ as a function of time $n$ for the Floquet model of MBL with the size $L=16$ and the disorder strength $W=6.2$, as well as the size $L=18$ and the disorder strength $W=6.4$. }\label{figr_t_floquet}
\end{figure}

In Fig.~\ref{figr_t_floquet}, we show the temporal fluctuation of the displacement $\text{Std}(\Delta X^{2})$ with different time $n$. Here, we consider two system sizes $L=16$ and $18$. We adopt $W= 6.2$ and $6.4$, near the maximum point of $\overline{\Delta X^{2}}$ [see Fig.~\ref{fig2_new}(g)], for $L=16$ and $18$, respectively.   It is seen that the saturation of $\text{Std}(\Delta X^{2})$ occurs at much earlier times than the Heisenberg time. Specifically, for $L=16$ and $18$, the chosen time interval $n\in[10^{4},2\times 10^{4}]$ and $n\in[2\times10^{4},3\times 10^{4}]$ lie in the the saturation region of $\text{Std}(\Delta X^{2})$.

%fig_std_t_floquet

\section{Details of data collapse}

\begin{figure*}[]
	\centering
	\includegraphics[width=1\linewidth]{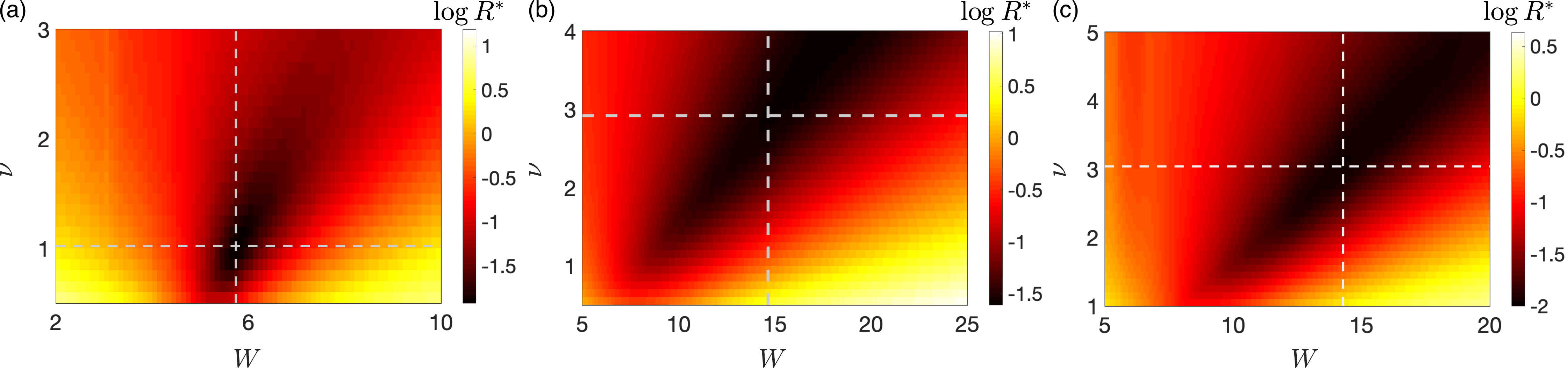}\\
	\caption{(a) The value of $\log R^{*}$ as a function of $W$  and $\nu$ for the Heisenberg model with quasiperiodic potential and $\lambda = 1.7$. (b) is similar to (a) but for the Heisenberg model with random potential and $\lambda = 1.2$. (c) The value of $\log R^{*}$ as a function of $W$  and $\nu$ for the Floquet model with $\lambda = 1.1$. The dashed lines represent the values of $\nu$ and $W$ with the minimum of $\log R^{*}$.}\label{figr_dc}
\end{figure*}

 \begin{figure}[]
	\centering
	\includegraphics[width=1\linewidth]{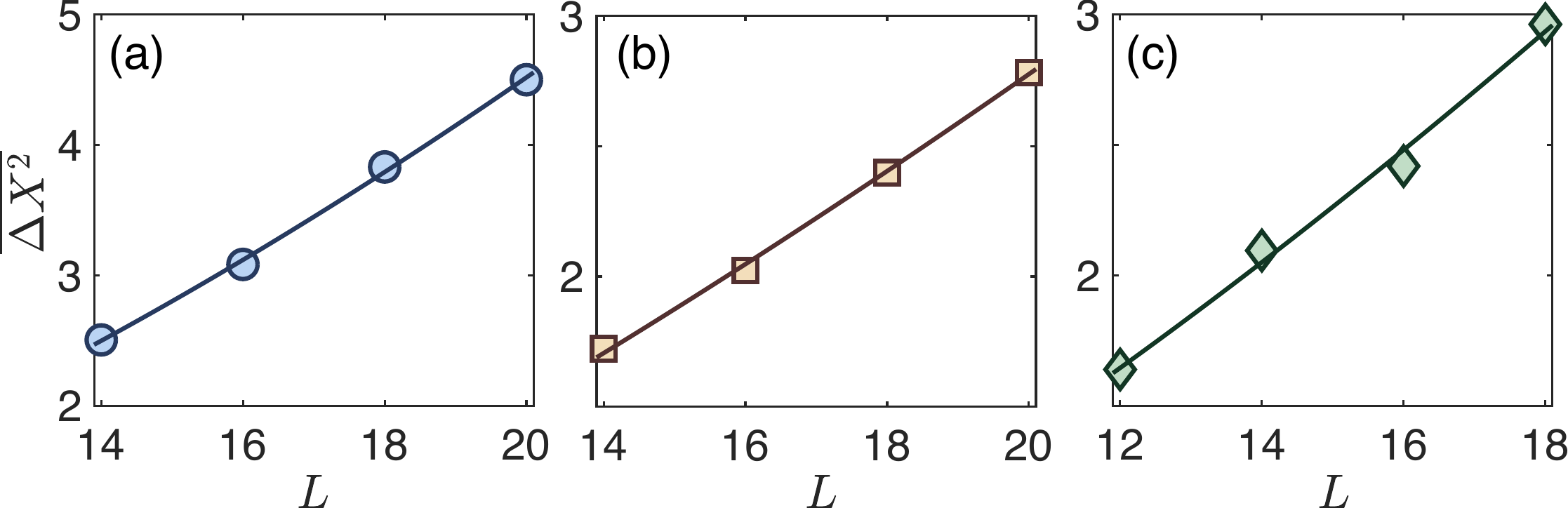}\\
	\caption{(a) For the quasiperiodic model with $W=5.7$, the time-averaged displacement $\overline{\Delta X^{2}}$ with the time interval $t\in[t_{H},t_{H}+100]$ ($t_{H}$ is the Heisenberg time) as a function of the system size $L$.  (b) is similar to (a) but for the random model with $W=15$.  (c)  For the Floquet model with $W=14.3$, the time-averaged displacement $\overline{\Delta X^{2}}$ as a function of the system size $L$. Here, for the data with $L=12$, $14$, and $16$, the time interval is $n\in[10^{4},2\times 10^{4}]$. For the data with $L=18$, the time interval is $n\in[3\times 10^{4},4\times 10^{4}]$. The solid lines are the power-law fitting $\overline{\Delta X^{2}} \propto L^{\beta}$. With 95\% confidence bounds,  the fitting gives $\beta\in[1.47,1.86]$, $\beta\in[1.20,1.53]$, and $\beta\in[1.04,1.81]$ for the quasiperiodic model, random model, and Floquet model, respectively.)}\label{figr_fit_add}
\end{figure}

In this section, we present the details of data collapse based on both the power-law and BKT-type correlation lengths. We first consider the data collapse based on the power-law correlation length (\ref{scaling}).  The method of performing data collapse is based on Ref.~\cite{Bhattacharjee:2001ux}. Finite size scaling has become an fundamental framework for characterizing critical properties of phase transition. A quantity $Q(t, L)$ with scaling depending on two variables, $t$ and $L$ can be expressed as
\begin{equation}
    Q(L, t) = L^\lambda f(t \cdot L^{1/\nu}).
\end{equation}
Data collapse is a quantitative way to show scaling and determine the exponents and the scaling function $f(x)$ that shows the critical behavior. In this work, we assume that the quantity takes the form of scaling function $Q = \overline{\Delta X^2}(L, W)=L^\lambda g\left(\left(W-W_c\right) L^{1 / \nu}\right)$. Therefore, the mean residual over all the data points $R=\sum\left|Q-L^{\lambda} g\left(\left(W-W_c\right) L^{1 / \nu}\right)\right|/N$ ($N$ is the number of data points) reaches minimum for the right choice of $(W_c, \lambda, \nu)$.

However, we cannot know the exact function form of $g(x)$ with finite known data points. Instead, we estimate the function values using an interpolation scheme, since the function can generally be assumed as an analytic function. Then we can define $R^*$ as
\begin{equation}
    R^*=\frac{1}{N} \sum_i \sum_{i \neq j} \sum_{n}\left|Q_{n, j} - L_j^{\lambda}\mathcal{G}_i\left(\left(W_{n,j}-W_c\right) L_j^{1 / \nu}\right)\right|,
    \label{r_eq}
\end{equation}
where $(Q_{n,j}, W_{n,j})$ denotes the $n$-th data point of $\overline{\Delta X^2}$ and the disorder strength $W$ for the system size $L_j$, and $\mathcal{G}_i$ is the interpolating function based on the data points $(Q_{n,i}, W_{n,i})$ for the system size $L_i$.  Here, we use the linear interpolation for $\mathcal{G}_i$. An optimization process can then be performed to obtain the minimum value of $R^*$, determining the values of $(W_c, \lambda, \nu)$.

For the Heisenberg model with random and quasiperiodic potential, the best data collapse is achieved for the parameters $(W_c, \lambda, \nu)=(5.7,1.7,1.0)$ and $(14.7,1.2,2.9)$, respectively. For the Floquet model, the parameters obtained from data collapse are $(W_c, \lambda, \nu)=(14.3,1.1,3.0)$. In Fig.~\ref{figr_dc}, we plot the value of $\log R^{*}$ as a function of $W$ and $\nu$, with a fixed value of $\lambda$. For the Heisenberg model with quasiperiodic and random fields, we chose $\lambda = 1.7$ and $\lambda = 1.2$, respectively, and for the Floquet model, the chosen value of $\lambda$ is $\lambda = 1.1$. 

Next, we estimate the errors of the data collapse, which can be obtained from the width of the minimum of $\log R^{*}$. Here, $\log R^{*}$ is a function of $W$, $\nu$, and $\lambda$, and the minimum point of $\log R^{*}$ is denoted as $\log R^{*}(W_{c},\nu_{0},\lambda_{0})$. The width of $W_{c}$ can be estimated as~\cite{Bhattacharjee:2001ux}
\begin{equation}
\Delta W = \eta W_{c} [2\log\frac{R^{*}(W_{c}+\eta W_{c},\nu_{0},\lambda_{0})}{R^{*}(W_{c},\nu_{0},\lambda_{0})}]^{-1/2}.
\label{error}
\end{equation}
with $\eta = 1\%$ being a chosen value. Similar to Eq.~(\ref{error}), the $\Delta \nu$ and $\Delta \lambda$ can also be estimated. For the Heisenberg model with random potential, $\Delta W \simeq 1.8$, $\Delta \nu \simeq 0.3$, and $\Delta \lambda \simeq 0.2$. For the Heisenberg model with quasiperiodic potential, $\Delta W \simeq 0.4$, $\Delta \nu \simeq 0.1$, and $\Delta \lambda \simeq 0.1$. For the Floquet model, $\Delta W \simeq 1.7$, $\Delta \nu \simeq 0.7$, and $\Delta \lambda \simeq 0.1$. 

Similar to Eq.~(\ref{r_eq}), for the BKT scaling, the cost function can be defined as 
\begin{equation}
    R^*=\frac{1}{N} \sum_i \sum_{i \neq j} \sum_{n}\left|Q_{n, j} - \mathcal{G}_{i} (L_{j} / \xi^{(n,j)}_{\text{BKT}})\right|
    \label{r_eq_bkt}
\end{equation}
with $\xi^{(n,j)}_{\text{BKT}} = \exp(b/\sqrt{W_{n,j} - W^{*}})$. The errors of the data collapse can also be estimated by Eq.~(\ref{error}). 

\section{Quality of power-law scaling}

In Appendix D, we analyze the the errors of the data collapse based on Eq.~(\ref{error}). It is seen that the values of $\Delta W$ for the random model and the Floquet model are significantly larger than that of the quasiperiodic model, indicating that the quality of power-law scaling for the quasiperiodic model is more compelling than other models. In this section, we adopt another method to quantify the quality of power-law scaling. 

Here, we fit the data of $\overline{\Delta X^{2}}$ as a function of the system size $L$, around the estimated critical point $W_{c}$ via the power-law scaling. We consider the fitting function $\overline{\Delta X^{2}}\propto L^{\beta}$. Near the estimated critical point $W_{c}$, if the value of $\beta$ obtained by the power-law fitting is close to the $\lambda$ estimated by the data collapse based on Eq.~(\ref{scaling}), the scaling analysis is reliable.

The results are displayed in Fig.~\ref{figr_fit_add}. We find that for the MBLT in the quasiperiodic model, the value of $\beta$ obtained by the power-law fitting is well consistent with the $\lambda$ extracted from the power-law scaling analysis [see Fig.~\ref{fig2_new}(b)].   However, as shown in Fig.~\ref{figr_fit_add}(b) and (c), the extracted values of $\beta$ for the MBLTs in the random-field Heisenberg model and the Floquet model deviate significantly from the corresponding $\lambda$ values. This discrepancy indicates that the power-law scaling fails to accurately describe the MBLT in these models. Instead, our revised analysis suggests that the BKT- type scaling provides a more appropriate framework for characterizing the critical behavior in these cases.

\bibliography{referencembl}

\end{document}